\documentclass[apjl]{emulateapj}

\usepackage{natbib}
\usepackage{color}

\newcommand{\myemail}{humehata@eso.org}

\slugcomment{draft September 19, 2015}

\shorttitle{ Starbursts in a node of the cosmic web at z=3.1}
\shortauthors{Umehata et al.}

\begin{document}


\title{ALMA Deep Field in SSA22: A concentration of dusty starbursts in a z=3.09 protocluster core}


\author{H. Umehata\altaffilmark{1,2}, Y. Tamura\altaffilmark{2}, K. Kohno\altaffilmark{2,3}, R.J. Ivison\altaffilmark{1,4}, D.M. Alexander\altaffilmark{5}, J. Geach\altaffilmark{6}, B. Hatsukade\altaffilmark{7}, D.H. Hughes\altaffilmark{14}, S. Ikarashi\altaffilmark{8}, Y. Kato\altaffilmark{7,15}, T. Izumi\altaffilmark{2}, R. Kawabe\altaffilmark{7,9,10}, M. Kubo\altaffilmark{11}, M. Lee\altaffilmark{7,15}, B. Lehmer\altaffilmark{12}, R. Makiya\altaffilmark{2}, Y. Matsuda\altaffilmark{7,9}, K. Nakanishi\altaffilmark{7,9}, T. Saito\altaffilmark{7}, I. Smail\altaffilmark{5}, T. Yamada\altaffilmark{13}, Y. Yamaguchi\altaffilmark{2}, M. Yun\altaffilmark{14}}
\affil{$^1$ European Southern Observatory, Karl-Schwarzschild-Str. 2, D-85748 Garching, Germany; \textcolor{blue}{{\myemail}}
\\
$^2$ Institute of Astronomy, School of Science, The University of Tokyo, 2-21-1 Osawa, Mitaka, Tokyo 181-0015, Japan
\\
$^3$ Research Center for the Early Universe, The University of Tokyo, 7-3-1 Hongo, Bunkyo, Tokyo 113-0033
\\
$^4$ Institute for Astronomy, University of Edinburgh, Royal Observatory, Blackford Hill, Edinburgh EH9 3HJ, UK
\\
$^5$ Centre for Extragalactic Astronomy, Department of Physics, Durham University, South Road, Durham, DH1 3LE, UK
\\
$^6$ Centre for Astrophysics Research, Science \& Technology Research Institute, University of Hertfordshire, Hatfield AL10 9AB, UK
\\
$^7$ National Astronomical Observatory of Japan, 2-21-1 Osawa, Mitaka, Tokyo 181-8588, Japan
\\
$^8$ Kapteyn Astronomical Institute, University of Groningen, P.O. Box 800, 9700AV Groningen, The Netherlands
\\
$^9$ Department of Astronomy, School of Science, The Graduate
University for Advanced Studies (SOKENDAI), Osawa, Mitaka,
Tokyo 181-8588, Japan
\\
$^{10}$ Joint ALMA Observatory, Alonso de Cordova 3107 Vitacura,
Santiago 763 0355, Chile
\\
$^{11}$ Institute for Cosmic Ray Research, University of Tokyo, 5-1-5 Kashiwa-no-Ha, Kashiwa City, Chiba 277-8582, Japan
\\
$^{12}$ Department of Physics, University of Arkansas, 226 Physics Building, 835 West Dickson Street, Fayetteville, AR 72701, USA
\\
$^{13}$ Astronomical Institute, Tohoku University, 6-3 Aoba, Aramaki, Aoba-ku, Sendai, Miyagi 980-8578, Japan
\\
$^{14}$ Department of astronomy, University of Massachusetts, Amherst, MA 01003, USA
\\
$^{15}$Department of Astronomy, Graduate school of Science, The University of Tokyo, 7-3-1 Hongo, Bunkyo-ku, Tokyo 133-0033, Japan
}

%




\begin{abstract}
We report the results of $1^{\prime}.5 \times3^{\prime}$ mapping at 1.1~mm with the Atacama Large Millimeter/submillimeter Array (ALMA) toward the central region of the $z=3.09$ SSA22 protocluster.
By combining our source catalog with archival spectroscopic redshifts, we find that eight submillimeter galaxies (SMGs) with flux densities, $S_{\rm 1.1~mm}=0.7-6.4$~mJy  ($L_{\rm IR}\sim10^{12.1}-10^{13.1}L_\odot$) are at $z=3.08-3.10$.
Not only are these SMGs members of the protocluster but they in fact reside within the node at the junction of the 50 Mpc-scale filamentary three-dimensional structure traced by Lyman-$\alpha$ emitters (LAEs) in this field. 
The eight SMGs account for a star formation rate density (SFRD) $\sim$10 $M_\odot$ yr$^{-1}$ Mpc$^{-3}$ in the node, which is two orders of magnitudes higher than the global SFRD at this redshift.
We find that four of the eight SMGs host a X-ray luminous active galactic nuclei (AGN). 
Our results suggest that the vigorous star formation activity and the growth of super massive black holes (SMBHs) occurred simultaneously in the densest regions at $z\sim3$, which may correspond to the most active historical phase of the massive galaxy population found in the core of the clusters in the present universe.
Two SMGs are associated with Lyman-$\alpha$ blobs (LABs), implying that
the two populations coexist in high density environments for a few cases.
\end{abstract}


\keywords{galaxies: starburst -- galaxies: quasars -- cosmology: large-scale structure of universe}



\section{Introduction}

In the current framework of cold dark matter (CDM) cosmologies, the formation and evolution of galaxies and super massive black holes (SMBHs) are closely related to those of cosmic structures on a large-scale. The density distribution of dark matter is expected to reflect that of baryonic matter and consequently that of galaxies (e.g., \citealt{1999MNRAS.307..529K}).
Therefore the environment where galaxies inhabit is a crucial key to comprehend galaxy formation and evolution throughout cosmic history.
In the local universe, the dense fields, seen as galaxy clusters, are occupied by passive, early-type galaxies while star-forming, late-type galaxies are seen in less dense fields (e.g., \citealt{1980ApJ...236..351D}).
Some works argue that the dependence of star formation rate (SFR) to density can be reversed at $z\gtrsim$1 (e.g., \citealt{2007A&A...468...33E}).
There are contrary claims(e.g., \citealt{2011MNRAS.418..938G}), although these are not based on dust-insenstitive tracers of star formation.

Submillimeter galaxies (SMGs) (for a recent review, \citealt{2014PhR...541...45C}) are one of the most important populations in unveiling the environmental dependence of galaxy formation on a large-scale in the early universe ($z\gtrsim2-3$).
 SMGs are massive gas-rich galaxies characterized as being enshrouded by dust and undergoing intense starburst activity (SFR of $\sim$ several 100-1000 M$_\odot$ yr$^{-1}$, e.g., \citealt{2014MNRAS.438.1267S}).
 A fraction of SMGs harbor active galactic nuclei (AGN), which suggests that SMGs also exist at the growth phase of SMBHs (e.g. \citealt{2005ApJ...632..736A}).
It has been argued that SMGs are progenitors of massive elliptical galaxies in the local universe (e.g., \citealt{1999ApJ...515..518E}), and cosmological simulations suggest the growth of the massive ellipticals in high density regions at $z\gtrsim2-3$ 
(e.g., \citealt{2006MNRAS.366..499D}).
Thus, unveiling the relationship between SMGs and underlying large-scale structures is quite important to understand the formation history of massive galaxies and large-scale structures. 

The SSA22 protocluster at $z=3.09$, which is considered to be an ancestor of present day clusters such as Coma (\citealt{1998ApJ...492..428S}), is one of the best field from such a viewpoint.
\citet{2012AJ....143...79Y} conducted a huge narrow band survey to detect $z\sim3.09$ Lyman-$\alpha$ emitters (LAEs) in over 2 deg$^2$ area and found that the density peak in SSA22 is extremely rare and outstanding ($\sim$6 times the average surface density).
Other populations such as Lyman break galaxies (LBGs, e.g., \citealt{1998ApJ...492..428S}) and distant red galaxies (DRGs, e.g., \citealt{2012ApJ...750..116U}) are also overabundant in this field.
While some submm/mm surveys taken with AzTEC/ASTE (\citealt{2009Natur.459...61T}; \citealt{2014MNRAS.440.3462U}), SCUBA (e.g., \citealt{2005MNRAS.363.1398G}; \citealt{2005ApJ...622..772C}) and SCUBA2 (\citealt{2014ApJ...793...22G}) have been conducted in this field, 
the angular resolution of single-dish telescopes is insufficient 
in obtaining an accurate picture.
We performed wide and deep imaging at 1.1~mm using the Atacama Large Millimeter/submillimeter Array (ALMA)  in this field in order to take advantage of its high angular resolution and high sensitivity to conduct an unconfused survey of dusty galaxy activity in this structure and pinpoint the galaxies responsible for it.

In this letter, we present the first results focusing on the SMGs at $z=3.09$.
The survey design and source catalog of the ALMA observations will be described in more detail in an upcoming paper (Umehata et al. in preparation).
Our observations are briefly explained in \S 2. We present the relevant  results in \S 3 and discuss the role of the environment in galaxy formation in \S 4.
Throughout this letter, we adopt a cosmology with 
$\Omega_{\rm m}=0.3, \Omega_\Lambda=0.7$, and H$_0$=70 km s$^{-1}$ Mpc$^{-1}$.

\section{Observations and Data Reduction}

In ALMA Cycle 2, we observed a $2^\prime \times 3^\prime$ area centered at RA (J2000) = 22$^{\rm h}$17$^{\rm m}$34$^{\rm s}$, Dec (J2000) = $+00^\circ17^{\prime}00^{\prime\prime}$ using 103 discrete pointing fields (Proposal ID 2013.1.00162.S, PI: Umehata).
We name this field ALMA Deep Field in SSA22 or ADF22.
In this letter we report the initial results from 80 pointing data, which roughly correspond to an area of $1.5^\prime \times 3^\prime$.
The observations were carried out during parts of five contiguous nights (6--10 June 2014) using 33--36 12m anntenas.
The array configuration was C34-4, which results in baseline lengths of 20--450~m.
We utilized the band 6 receiver with the TDM correlator to select a central frequency of 263~GHz (1.1~mm).
The on-source time per pointing is 2.0--2.5 min.
The quasar J2148+0657 was observed for bandpass, amplitude, phase, and flux calibration.

The data were processed with the Common Astronomy Software Application ({\sc casa}) \footnote{http://casa.nrao.edu}.
The final entire map was created through the 'clean' process (with natural weighting and by setting the imager mode to 'mosaic' in {\sc casa}).
The resulting map has a synthesized beam of $0.^{\prime\prime}53\times0.^{\prime\prime}50$ (P.A. $= -84$ deg) and a typical rms revel of 0.07 mJy beam$^{-1}$.
We utilize a source-finding code, {\sc aegean} v1.9.5-56 (\citealt{2012MNRAS.422.1812H}), to extract sources on the final mosaic image with a detection threshold of 4$\sigma$.
The flux densities of the detected sources were measured with {\sc casa} task, {\sc imfit}.

\section{Results}

\subsection{Extraction of $z=3.09$ SMGs}

\begin{figure}
\epsscale{1.15}
\plotone{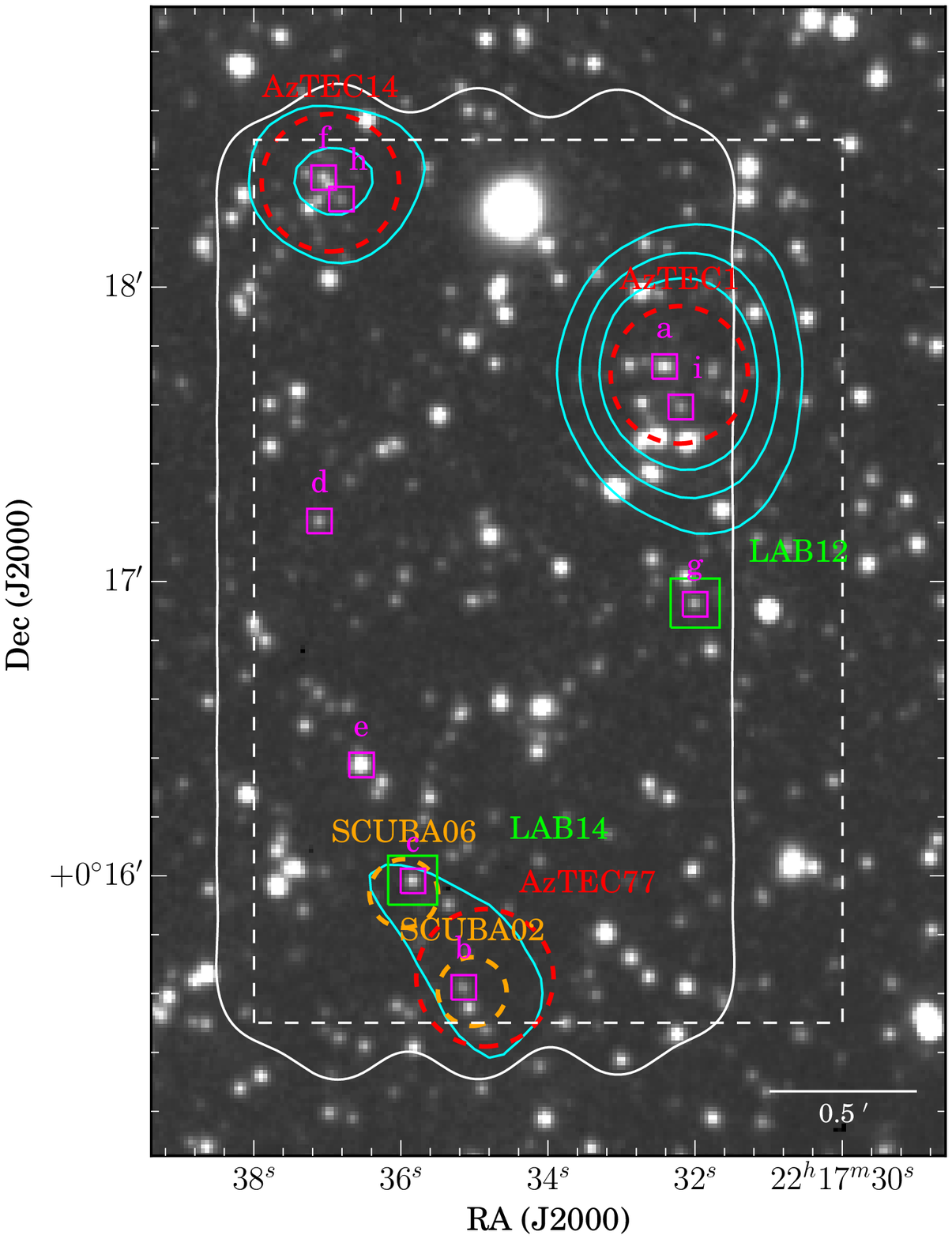}
\caption{
Spatial distributions of SMGs as well as LABs on the IRAC 4.5 $\mu$m image.
The 5$^{\prime\prime}\times5^{\prime\prime}$ magenta squares represent ALMA sources at $z=3.09$.
The red and orange dashed circles stand for AzTEC/ASTE (\citealt{2014MNRAS.440.3462U}) and SCUBA sources (\citealt{2006MNRAS.370.1057S}), which are suggested to be at $z\sim3.09$ (\citealt{2005ApJ...622..772C}; \citealt{2010ApJ...724.1270T}; \citealt{2014MNRAS.440.3462U}).
The diameters of these circles are 28$^{\prime\prime}$ and 14$^{\prime\prime}$ respectively, which correspond to the size of FWHM of these surveys.
The cyan contours show 3.0, 6.0, and 9.0 $\sigma$ of 1.1~mm emission in the AzTEC map.
The dashed white rectangle shows the whole 2$^\prime \times 3^\prime$ area of ADF22 and the white contours outline the currently obtained area (the primary beam attenuation is less than 50\%).
The green boxes show the position of the LABs (\citealt{2004AJ....128..569M}).
}
\label{f1}
\end{figure}

In order to determine redshifts of ALMA sources and extract the members of the $z=3.09$ structure,
we compared our 1.1~mm source catalog with archival catalogs of spectroscopic and photometric redshifts (\citealt{2013MNRAS.429.3047B}; \citealt{2010ApJ...724.1270T}; \citealt{2015ApJ...799...38K};  \citealt{2005ApJ...622..772C}; Yun et al., in preparation).
As a result, we yielded eight 1.1~mm sources with $z_{\rm spec}=3.08-3.10$ and a further SMG with $z_{\rm phot}\sim3.1$.
The positions of these ALMA-selected SMGs are shown in Fig. \ref{f1}.

The archival spectroscopic redshifts are measured from mm, near-infrared, and optical spectroscopy.
Millimeter spectroscopy of $^{12}$CO (3-2) reveals the redshifts for two SMGs.
ADF22b is consistent with SMM J221735.15+001537.2 at $z_{\rm CO}=3.096$, which were observed with PdBI (\citealt{2013MNRAS.429.3047B}).
In addition, one of the SMGs discovered in the AzTEC/ASTE survey, SSA22-AzTEC1 (hereafter AzTEC1, \citealt{2014MNRAS.440.3462U}), has been recently observed with the Large Millimeter Telescope (LMT) with an effective beam size of $\sim28^{\prime\prime}$.
Its redshift is $z=3.092$ if we consider that the detected line is $^{12}$CO (3-2) (Yun et al. in preparation). 
We find that AzTEC1 is split into two sources of ADF22a and ADF22i in our ALMA map.
ADF22a has a mid-infrared to radio photo-$z$ of $3.19^{+0.26}_{-0.35}$ (\citealt{2010ApJ...724.1270T}) and likely dominates the CO emission since it is $3\times$ brighter than ADF22i at 1.1 mm.  Thus we conclude that at least ADF22a is at $z=3.092$.
For ADF22i, we derive an optical to near-infrared photo-$z$ of 3.08$^{+0.17}_{-0.15}$ using a method in \citet{2014MNRAS.440.3462U}. 

\begin{figure}
\epsscale{1.15}
\plotone{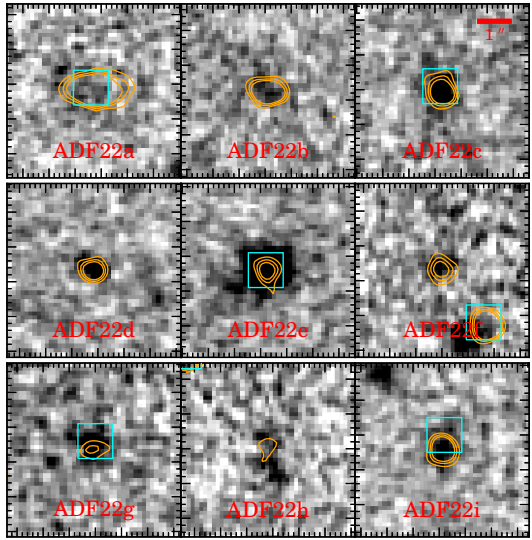}
\caption{
ALMA 1.1mm contours of $z\simeq3.09$ SMGs overlayed on the MOIRCS {\it Ks}-band image.
Contours of 3, 6, and 9$\sigma$ are shown.
Cyan squares represent the position of X-ray sources (\citealt{2009MNRAS.400..299L}).
Each panel is 5$^{\prime\prime}$ square with the ALMA position centered.
All the SMGs apart from ADF22a have $Ks$-band counterparts.
}
\label{f2}
\end{figure}

The spectroscopic redshifts of five SMGs (ADF22d, ADF22e, ADF22f, ADF22g, ADF22h) are determined using near-infrared spectroscopy performed with Subaru/MOIRCS.
\citet{2015ApJ...799...38K} reported the detection of a [OIII]$\lambda$5007 line for the {\it Ks}-band counterparts of the five SMGs.
We also confirmed that ADF22c coincides with a radio source at $z_{\rm Ly\alpha}$=3.089, which has been considered as the most plausible counterpart of SCUBA06 (SMM J221735.84+001558.9, \citealt{2005ApJ...622..772C}).
We present the positional relationship between the SMGs and {\it Ks}-band counterparts in Fig. \ref{f2}.
The synthesized beam of our observation and 4$\sigma$ detection threshold yield an expected astrometric accuracy of $\lesssim0^{\prime\prime}$.15 (e.g., \citealt{2013ApJ...776...22H}).
All SMGs except for ADF22a has {\it Ks}-band counterparts within 0$^{\prime\prime}$.18 (or $\approx$1.4 kpc at $z\approx3.09$).
Therefore we concluded that the [OIII] and Ly$\alpha$ lines are from the SMGs.

While some of the SMGs harbor a X-ray AGN (\S 3.2), which can substantially contribute the energetic output at shorter wavelengths (e.g., \citealt{2008ApJ...684..136G}), $S_{\rm 1.1~mm}$ is considered not significantly contaminated by AGNs and hence can be a good tracer of star-formation activity.
It is known that AGN contribution falls steeply at restframe $\gtrsim$40 $\mu$m (e.g., \citealt{2011MNRAS.414.1082M}), while we observe  at restframe $\approx$279 $\mu$m.
We estimated the potential AGN contribution to $S_{\rm 1.1~mm}$ using MIPS 24 $\mu$m flux densities ($S_{24\mu \rm{m}}$), following Alexander et al. in preparation.
ADF22e is the brightest at 24 $\mu$m ($S_{24\mu \rm{m}}$=450$\pm$10 $\mu$Jy, e.g., \citealt{2009ApJ...692.1561W}).
Assuming conservatively that $S_{24\mu \rm{m}}$  is fully powered by AGN, 
the predicted $S_{\rm 1.1~mm}$ based on the averaged empirical AGN spectral energy
distribution (SED) template of \citet{2011MNRAS.414.1082M} is $S_{1.1\rm{mm}}\sim110$ $\mu$Jy (or $\sim$11\% of the observed flux density).
The remaining X-ray SMGs are relatively faint at 24 $\mu$m ($\lesssim100$ $\mu$Jy, e.g., \citealt{2009ApJ...692.1561W}), which corresponds to $S_{\rm 1.1~mm}$ $\lesssim$ 25 $\mu$Jy.
Therefore the implied AGN contribution to $S_{\rm 1.1~mm}$ is $\lesssim 3.5$\%.

The lack of any other submm/mm data with comparable angular resolution prevents from putting a constraint on dust SEDs. To evaluate SFR of the SMGs, we calculated the infrared luminosities ($L_{\rm IR}$ [8-1000 $\mu$m]) using SED templates of well-studied starburst galaxies, Arp 220 and M82 (GRASIL; \citealt{1998ApJ...509..103S}), a composite SED of SMGs from the ALESS (\citealt{2014MNRAS.438.1267S}), and SMMJ2135--0201 (the cosmic eyelash; \citealt{2010Natur.464..733S}) to consider a variety of SEDs.
We created best-fit SED for each template based on redshift and $S_{\rm 1.1~mm}$.
The spectra between 8 and 1000 $\mu$m in the restframe were integrated and we derive a median value as well as minimum/maximum values.
We convert $L_{\rm IR}$ to the SFR using SFR/$M_\odot$yr$^{-1}$ = 1.0$\times10^{-10}$ $L_{\rm IR}$/$L_\odot$ (\citealt{1998ARA&A..36..189K}), assuming a Chabrier initial mass function (\citealt{2003PASP..115..763C}).

\subsection{SMGs in the node of the Cosmic Web}

\begin{figure}
\epsscale{0.95}
\plotone{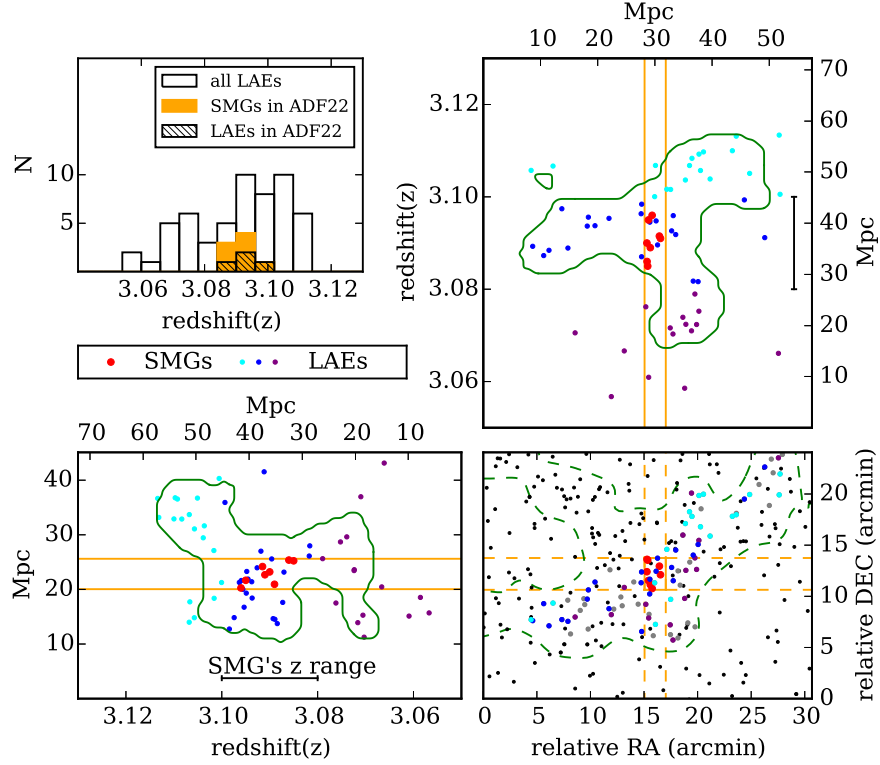}
\caption{
({\it top left})
Redshift distributions of eight SMGs and four LAEs in ADF22 overlayed on that of 56 LAEs in the SSA22 field (\citealt{2005ApJ...634L.125M}).
({\it top right, left and right bottom})
Three-dimensional distributions of SMGs in ADF22 (big red circles) and narrowband-selected LAEs in the SSA22 field (small colored circles; \citet{2005ApJ...634L.125M}). 
The solid (dashed) green lines show the projected contour of the local volume (surface) density of the LAEs.
For more detail, please see \citet{2005ApJ...634L.125M}.
The eight SMGs are concentrated in the intersection of the LAE filamentary structure.}
\label{f3}
\end{figure}

One of the distinguishing characteristics of the SSA22 protocluster at $z=3.09$ is the existence of a three-dimensional large-scale structure traced by LAEs (\citealt{2005ApJ...634L.125M}).
Matsuda et al. confirmed that 56 narrow-band selected LAE candidates were actually at $z=3.06-3.12$ and found that the protocluster seen as two-dimensional density excess of LAEs had 50 Mpc-scale three-dimensional filamentary structure (Fig. \ref{f3}).
ADF22 was designed to observe the intersection of the three-dimensional structure and the spectroscopic redshifts enable to compare the distribution of SMGs against the large-scale structure three-dimensionally.
All eight SMGs with spec-$z$ are distributed in a range of $z=3.08-3.10$, which is in good agreement with the redshift space of the node of the LAE filamentary structure as illustrated in Fig. \ref{f3}.
As noticed in \citet{2005ApJ...634L.125M}, the redshifts determined by Lyman-$\alpha$ emission lines contain uncertainties due to the peculiar velocities and resonant scatterings in the outflowing \textsc{Hi} gas.
However, the estimated redshift dispersion is predicted to be small ($\sigma_{z}\sim0.005$) and therefore the uncertainties are not supposed to matter in the comparison as a whole.
ADF22i, which has been selected based on its photo-$z$, might be also in the node.

To evaluate the overabundance of SMGs in the node quantitatively, we calculate the volume density of the 1.1~mm sources and compare it with the expected value in general fields based on the total IR luminosity.
\citet{2014MNRAS.438.1267S} derived the IR luminosity functions of ALMA-selected SMGs using results of the ALESS survey in the ECDF-S. They utilized optical to near-infrared photo-$z$'s (\citealt{2014ApJ...788..125S}) and IR luminosities derived from SED fitting to determine that the volume density of SMGs at $z=2.5-3.5$ is about $4\times10^{-6}$ Mpc$^{-3}$ for $L_{\rm IR}\gtrsim10^{12.5}L_\odot$ \footnote{
We note that some works suggested that ECDF-S could be $\sim \times2$ underdense compared to other submm surveys (e.g., \citealt{2014MNRAS.438.1267S}). 
}.
Assuming a redshift slice of $z=3.08-3.10$, 
the predicted number of SMGs is 1$\times10^{-3}$ within this 1.5$^\prime \times 3^\prime$ field, which corresponds to an area of 2.8 $\times$ 5.6 Mpc at $z=3.09$.
There are three SMGs with $z_{\rm spec}=3.08-3.10$ and $L_{\rm IR}\gtrsim 10^{12.5}L_\odot$ in ADF22 as listed in Table 1, which is 
three orders of magnitude greater than the number expected for general fields.
Thus the volume density of SMGs is unusually high in the node.
We calculate a star-formation rate density (SFRD) in the node of  $\sim$10 $M_\odot$ yr$^{-1}$ Mpc$^{-3}$ considering the eight SMGs with spec-$z$, which is two orders of magnitude higher than the global SFRD at this redshift (\citealt{2014ARA&A..52..415M}).

In addition to the overabundance, the $z=3.09$ SMGs in ADF22 are characterized by overlaps with AGNs and Lyman-$\alpha$ blobs (LABs), which are extended Lyman-$\alpha$ emitting nebulae.
The whole area of ADF22 was observed by $Chandra$ at 0.5--2 keV and 2--8 keV (\citealt{2009MNRAS.400..299L}).
We found that four SMGs at $z=3.09$ (ADF22a, ADF22c, ADF22e, ADF22g) have X-ray counterparts listed in \cite{2009ApJ...691..687L} within 0$^{\prime\prime}$.5, which means that 50$^{+39}_{-24}$\% of SMGs in the node host X-ray luminous AGNs with $L_X\sim10^{44}$ erg s$^{-1}$ (\citealt{2009ApJ...700....1G}, \citealt{2010ApJ...724.1270T}).
ADF22i also has a X-ray counterpart and hence it is one of such AGN-host SMGs.
The ALESS survey have comparable depth at submm/mm wavelengths compared to ADF22.
In the field, \citet{2013ApJ...778..179W} estimated an AGN fraction --a fraction of SMGs containing AGN-- of about $10$\% for AGNs with rest-frame 0.5--8.0 keV apparent luminosity $\gtrsim10 ^{43}$ erg s$^{-1}$, which is equivalent to that limit at $z=3.09$ in ADF22 (\citealt{2009MNRAS.400..299L}).
Thus the AGN fraction is relatively high compared with that typically found in the whole SMG population.
We also note that there is another X-ray AGN without ALMA detection (J221737.3+001823.2, \citealt{2015ApJ...799...38K}).
In ADF22, there are two of 35 LABs listed in \citet{2004AJ....128..569M}.
\citet{2009ApJ...700....1G} reported that X-ray counterparts of two SMGs (ADF22c, ADF22g) were associated
with the two LABs (LAB14, LAB12), respectively.
In other word,
two SMGs in the node are found to be associated with LABs.

\section{Discussion \& Conclusions}

There is vigorous debate on the environmental dependence on SMG formation at $z\gtrsim2$, the peak era of cosmic star formation activity.
Some studies show that a handful of SMGs at $z=4-5$ coexist with other star-forming galaxies such as LBGs, and argue that SMGs are at protoclusters (e.g., \citealt{2009ApJ...694.1517D}; \citealt{2011Natur.470..233C}; \citealt{2012Natur.486..233W}).
Recent submm/mm surveys performed with single-dish telescopes unveil the possible excess of SMGs at $z=2-3$ towards the known overdense fields (e.g., \citealt{2009Natur.459...61T}; \citealt{2014MNRAS.440.3462U}) or a serendipitously found protocluster (\citealt{2015arXiv150601715C}).
On the other hand, \citet{2009ApJ...691..560C} and \citet{2015MNRAS.452..878M} suggest that SMGs are poor tracers of massive protoclusters and they can reside in less dense environments at least at $z=2-2.5$.

There have been some obstacles to resolve this issue.
First, there is ambiguity in the definition of a protocluster in terms of protocluster mass and the degree of galaxy assembly.
Second, the angular resolution of a single-dish telescope is generally insufficient to identify counterparts in other wavelength images reliably.
Third, the \emph{evolution} of the relationship between SMGs and environment has not been characterized well, although the protocluster mass scale and star formation in member galaxies are expected to strongly evolve through the era in which SMGs have been observed ($1 \lesssim z \lesssim 6$, e.g., \citealt{2014PhR...541...45C}).

The $z=3.09$ SSA22 protocluster is a preferable target in unveiling whether SMGs are formed in high density environments at the epoch by virtue of having a remarkable large-scale structure. 
As described above, our results from ADF22 put significant an observational evidence on the site of SMG formation.
Our finding, overabundance of SMGs in ADF22, is in line with that of the previous works which claim that the associations of SMGs trace the dense environment.
Furthermore it might be suggested that SMGs are preferentially formed around the intersections of the cosmic filamentary structure, though we should note that our ALMA view is limited to only the central part.
The existence of a number of SMGs at the $z=3.09$ protocluster core is also suggestive in considering the evolution of the relationship between SMGs and (proto)clusters.
\citet{2014ApJ...782...19S} reported that SMGs are located at the less dense parts in a $z=1.6$ cluster mainly based on their SCUBA2 observations, and the very central area is occupied by the most massive red quiescent galaxies.
Meanwhile \citet{2015ApJ...806..257M} find excess of SCUBA2 sources in a cluster at $z=1.5$, suggesting a variety of environments associated with SMG activity at $z<2$. 
In contrast to such a result at a relatively low redshift, the SSA22 protocluster at $z=3.09$ seems to be in a phase in which the most intense starbursts occur at the core.
The overabundance of SCUBA sources around high redshift radio galaxies at $z\sim3$, which is considered to be sign posts of high density peaks, could show similar phases (e.g., \citealt{2000ApJ...542...27I}; \citealt{2003ApJ...583..551S}), though a shortage of angular resolution and a deficit of redshift information prevent further comparison.
These case studies indicate that we might be seeing the exact growth phase of stellar components as well as SMBHs of the massive ellipticals seen in the core of the present-day clusters, and the site where the most active populations reside would evolve from the center to outer parts at decreasing redshifts.

The other important aspect of the nature of SMGs in the node is a correlation between SMGs and other populations.
First, the SMGs in ADF22 are frequently associated with AGNs.
Our results suggest that the unique environment, the intersection of the large-scale structure, can lead to this trend.
Some ideas previously presented; both of accelerated infall of gas and a higher rate of mergers are naturally expected in overdense environments (e.g., \citealt{2006MNRAS.366..499D}), which might account for the  overabundance of dusty starbursts and their high AGN fraction.
Additionally, ALMA imaging helps us uncover obscured star-forming cores embedded in LABs.
Intense star formation activities and/or AGNs are supposed to be the possible origins of the extended Lyman-$\alpha$ emission (e.g., \citealt{2000ApJ...532L..13T}) and therefore a connection between SMGs and LABs has been considered.
We identify two X-ray luminous SMGs associated with LABs, which shows that star formation and/or AGNs in SMGs can be related to LABs at least for some cases.
On the other hands, the majority of SMGs do not seem to be accompanied by giant Lyman-$\alpha$ nebulae in their active starburst phase, while both populations inhabit high density environments.

\acknowledgments
We thank the anonymous referee gratefullly.
This paper makes use of the following ALMA data: ADS/JAO.ALMA\#2013.1.00162.S (PI: H. Umehata). ALMA is a partnership of ESO (representing its member states), NSF (USA) and NINS (Japan), together with NRC (Canada) and NSC and ASIAA (Taiwan) and KASI (Republic of Korea), in cooperation with the Republic of Chile. The Joint ALMA Observatory is operated by ESO, AUI/NRAO and NAOJ.
HU, YM were supported by the ALMA Japan Research Grant of NAOJ Chile Observatory, NAOJ-ALMA-0071, 0086. HU, ML are thankful for the JSPS fellowship.
YM, BH acknowledge support from JSPS KAKENHI Grant Number 20647268 and 15K17616.
RJI acknowledges support from ERC in the form of the Advanced Investigator Programme, 321302, COSMICISM.
IRS acknowledges support from STFC (ST/L00075x/1), the ERC Advanced Investigator program
DUSTYGAL 321334, and a Royal Society/Wolfson Merit
Award. 
This work is based in part on archival data obtained with the NASA Spitzer Space Telescope.



{\it Facilities:} \facility{ALMA}, \facility{Subaru(MOIRCS)}, \facility{CXO(ASIS-I)},\facility{Spitzer (IRAC)}.

\begin{center}
\begin{deluxetable}{ccccccccc}
\tabletypesize{\scriptsize}
\tablecaption{z=3.09 SMGs in ADF22}
\tablewidth{0pt}
\tablehead{
\colhead{ADF22 ID} &\colhead{NAME} & \colhead{$z$} & \colhead{Type} & $S_{\rm peak}$/N& $S_{\rm 1.1 \ mm}$ & log ($L_{\rm IR}$) & SFR$_{\rm IR} $& X-ray AGN  \\
& &  & &  & (mJy) & ( $L_{\odot}$) &(M$_\odot$ yr$^{-1}$)  
}
\startdata
ADF22a  &   ALMAJ221732.41+001743.8  &   3.092 &   $^{12}$CO(3-2) $^{(1)}$ & 38.8 &   6.4 $\pm$ 0.2 &     $ 13.1_{- 0.1 }^{+ 0.2 }$   &  1180$_{- 230 }^{+ 890 }$   & Y   \\   
ADF22b  &   ALMAJ221735.15+001537.3  &   3.096 &  $^{12}$CO(3-2) $^{(2)}$& 23.8 &   2.3 $\pm$ 0.1 &     $ 12.6_{- 0.1 }^{+ 0.2 }$  & 420$_{- 80 }^{+ 320 }$  & N    \\   
ADF22c  &   ALMAJ221735.83+001559.0  &   3.089 & Lyman-$\alpha$ $^{(3)}$& 16.6 &   1.8 $\pm$ 0.1 &    $ 12.5_{- 0.1 }^{+ 0.2 }$   &  330$_{- 50 }^{+ 250 }$   & Y   \\   
ADF22d  &   ALMAJ221737.11+001712.4  &   3.090 &  [OIII]$\lambda5007$ $^{(4)}$ & 15.2 &   1.1 $\pm$ 0.1 &     $ 12.3_{- 0.1 }^{+ 0.2 }$ & 200$_{- 40}^{+ 150 }$  & N    \\   
ADF22e  &   ALMAJ221736.54+001622.7  &   3.095 &  [OIII]$\lambda5007$ $^{(4)}$& 11.3 &   1.0 $\pm$ 0.1 &     $ 12.3_{- 0.1}^{+ 0.2 }$ & 180$_{- 35 }^{+ 140 }$   & Y    \\   
ADF22f  &   ALMAJ221737.05+001822.4  &   3.086 &  [OIII]$\lambda5007$ $^{(4)}$& 10.2 &   1.1 $\pm$ 0.1 &    $ 12.3_{- 0.1 }^{+ 0.2 }$  & 200$_{- 40 }^{+ 150 }$  & N    \\   
ADF22g  &   ALMAJ221732.01+001655.4  &   3.091 &  [OIII]$\lambda5007$ $^{(4)}$& 5.8 &   0.7 $\pm$ 0.1 &   $ 12.1_{- 0.1 }^{+ 0.2 }$  & 130$_{- 24 }^{+ 100 }$ & Y    \\   
ADF22h  &   ALMAJ221736.81+001818.1  &   3.085 &  [OIII]$\lambda5007$ $^{(4)}$& 4.3 &   0.8 $\pm$ 0.2 &   $ 12.2_{- 0.1 }^{+ 0.2 }$  &150$_{- 30 }^{+ 110 }$  & N    \\   
ADF22i  &   ALMAJ221732.19+001735.6  &   (3.08$^{+0.17}_{-0.15}$) &  phot-$z$ & 17.7 &   2.0 $\pm$ 0.1 &  $ 12.6_{- 0.1 }^{+ 0.2 }$  &370$_{- 70 }^{+ 280 }$  & Y    
\enddata
\tablecomments{
References of redshifts are: (1) Yun et al. in preparation, (2) \citet{2013MNRAS.429.3047B}, (3)  \citet{2005ApJ...622..772C}, (4) \citet{2015ApJ...799...38K}.
Photo-z of ADF22i is estimated in a similar way in \citet{2014MNRAS.440.3462U}.
For the column of X-ray AGN, Y means SMGs that host a X-ray luminous AGN (rest-frame 0.5--8.0 keV apparent luminosity $\gtrsim10 ^{43}$ erg s$^{-1}$, \citealt{2009MNRAS.400..299L}). N represent SMGs without a detectable X-ray counterpart.
}
\end{deluxetable}
\end{center}


\begin{thebibliography}{}
\expandafter\ifx\csname natexlab\endcsname\relax\def\natexlab#1{#1}\fi

\bibitem[{{Alexander} {et~al.}(2005){Alexander}, {Bauer}, {Chapman}, {Smail},
  {Blain}, {Brandt}, \& {Ivison}}]{2005ApJ...632..736A}
{Alexander}, D.~M., {Bauer}, F.~E., {Chapman}, S.~C., {et~al.} 2005, \apj, 632,
  736

\bibitem[{{Bothwell} {et~al.}(2013){Bothwell}, {Smail}, {Chapman}, {Genzel},
  {Ivison}, {Tacconi}, {Alaghband-Zadeh}, {Bertoldi}, {Blain}, {Casey}, {Cox},
  {Greve}, {Lutz}, {Neri}, {Omont}, \& {Swinbank}}]{2013MNRAS.429.3047B}
{Bothwell}, M.~S., {Smail}, I., {Chapman}, S.~C., {et~al.} 2013, \mnras, 429,
  3047
  
\bibitem[{{Capak} {et~al.}(2011){Capak}, {Riechers}, {Scoville}, {Carilli},
  {Cox}, {Neri}, {Robertson}, {Salvato}, {Schinnerer}, {Yan}, {Wilson}, {Yun},
  {Civano}, {Elvis}, {Karim}, {Mobasher}, \& {Staguhn}}]{2011Natur.470..233C}
{Capak}, P.~L., {Riechers}, D., {Scoville}, N.~Z., {et~al.} 2011, \nat, 470,
  233

\bibitem[{{Casey} {et~al.}(2014){Casey}, {Narayanan}, \&
  {Cooray}}]{2014PhR...541...45C}
{Casey}, C.~M., {Narayanan}, D., \& {Cooray}, A. 2014, \physrep, 541, 45

\bibitem[{{Casey} {et~al.}(2015){Casey}, {Cooray}, {Capak}, {Fu}, {Kovac},
  {Lilly}, {Sanders}, {Scoville}, \& {Treister}}]{2015arXiv150601715C}
{Casey}, C.~M., {Cooray}, A., {Capak}, P., {et~al.} 2015, ArXiv e-prints,
  arXiv:1506.01715

\bibitem[{{Chabrier}(2003)}]{2003PASP..115..763C}
{Chabrier}, G. 2003, \pasp, 115, 763

\bibitem[{{Chapman} {et~al.}(2009){Chapman}, {Blain}, {Ibata}, {Ivison},
  {Smail}, \& {Morrison}}]{2009ApJ...691..560C}
{Chapman}, S.~C., {Blain}, A., {Ibata}, R., {et~al.} 2009, \apj, 691, 560

\bibitem[{{Chapman} {et~al.}(2005){Chapman}, {Blain}, {Smail}, \&
  {Ivison}}]{2005ApJ...622..772C}
{Chapman}, S.~C., {Blain}, A.~W., {Smail}, I., \& {Ivison}, R.~J. 2005, \apj,
  622, 772

\bibitem[{{Daddi} {et~al.}(2009){Daddi}, {Dannerbauer}, {Stern}, {Dickinson},
  {Morrison}, {Elbaz}, {Giavalisco}, {Mancini}, {Pope}, \&
  {Spinrad}}]{2009ApJ...694.1517D}
{Daddi}, E., {Dannerbauer}, H., {Stern}, D., {et~al.} 2009, \apj, 694, 1517

\bibitem[{{De Lucia} {et~al.}(2006){De Lucia}, {Springel}, {White}, {Croton},
  \& {Kauffmann}}]{2006MNRAS.366..499D}
{De Lucia}, G., {Springel}, V., {White}, S.~D.~M., {Croton}, D., \&
  {Kauffmann}, G. 2006, \mnras, 366, 499

\bibitem[{{Dressler}(1980)}]{1980ApJ...236..351D}
{Dressler}, A. 1980, \apj, 236, 351

\bibitem[{{Eales} {et~al.}(1999){Eales}, {Lilly}, {Gear}, {Dunne}, {Bond},
  {Hammer}, {Le F{\`e}vre}, \& {Crampton}}]{1999ApJ...515..518E}
{Eales}, S., {Lilly}, S., {Gear}, W., {et~al.} 1999, \apj, 515, 518

\bibitem[{{Elbaz} {et~al.}(2007){Elbaz}, {Daddi}, {Le Borgne}, {Dickinson},
  {Alexander}, {Chary}, {Starck}, {Brandt}, {Kitzbichler}, {MacDonald},
  {Nonino}, {Popesso}, {Stern}, \& {Vanzella}}]{2007A&A...468...33E}
{Elbaz}, D., {Daddi}, E., {Le Borgne}, D., {et~al.} 2007, \aap, 468, 33

\bibitem[{{Geach} {et~al.}(2005){Geach}, {Matsuda}, {Smail}, {Chapman},
  {Yamada}, {Ivison}, {Hayashino}, {Ohta}, {Shioya}, \&
  {Taniguchi}}]{2005MNRAS.363.1398G}
{Geach}, J.~E., {Matsuda}, Y., {Smail}, I., {et~al.} 2005, \mnras, 363, 1398

\bibitem[{{Geach} {et~al.}(2009){Geach}, {Alexander}, {Lehmer}, {Smail},
  {Matsuda}, {Chapman}, {Scharf}, {Ivison}, {Volonteri}, {Yamada}, {Blain},
  {Bower}, {Bauer}, \& {Basu-Zych}}]{2009ApJ...700....1G}
{Geach}, J.~E., {Alexander}, D.~M., {Lehmer}, B.~D., {et~al.} 2009, \apj, 700,
  1

\bibitem[{{Geach} {et~al.}(2014){Geach}, {Bower}, {Alexander}, {Blain},
  {Bremer}, {Chapin}, {Chapman}, {Clements}, {Coppin}, {Dunlop}, {Farrah},
  {Jenness}, {Koprowski}, {Micha{\l}owski}, {Robson}, {Scott}, {Smith},
  {Spaans}, {Swinbank}, \& {van der Werf}}]{2014ApJ...793...22G}
{Geach}, J.~E., {Bower}, R.~G., {Alexander}, D.~M., {et~al.} 2014, \apj, 793,
  22
  
  \bibitem[{{Gruppioni} {et~al.}(2008){Gruppioni}, {Pozzi}, {Polletta},
  {Zamorani}, {La Franca}, {Sacchi}, {Comastri}, {Pozzetti}, {Vignali},
  {Lonsdale}, {Rowan-Robinson}, {Surace}, {Shupe}, {Fang}, {Matute}, \&
  {Berta}}]{2008ApJ...684..136G}
{Gruppioni}, C., {Pozzi}, F., {Polletta}, M., {et~al.} 2008, \apj, 684, 136

\bibitem[{{Gr{\"u}tzbauch} {et~al.}(2011){Gr{\"u}tzbauch}, {Conselice},
  {Bauer}, {Bluck}, {Chuter}, {Buitrago}, {Mortlock}, {Weinzirl}, \&
  {Jogee}}]{2011MNRAS.418..938G}
{Gr{\"u}tzbauch}, R., {Conselice}, C.~J., {Bauer}, A.~E., {et~al.} 2011,
  \mnras, 418, 938

\bibitem[{{Hancock} {et~al.}(2012){Hancock}, {Murphy}, {Gaensler}, {Hopkins},
  \& {Curran}}]{2012MNRAS.422.1812H}
{Hancock}, P.~J., {Murphy}, T., {Gaensler}, B.~M., {Hopkins}, A., \& {Curran},
  J.~R. 2012, \mnras, 422, 1812

\bibitem[{{Hodge} {et~al.}(2013){Hodge}, {Carilli}, {Walter}, {Daddi}, \&
  {Riechers}}]{2013ApJ...776...22H}
{Hodge}, J.~A., {Carilli}, C.~L., {Walter}, F., {Daddi}, E., \& {Riechers}, D.
  2013, \apj, 776, 22

\bibitem[{{Ivison} {et~al.}(2000){Ivison}, {Dunlop}, {Smail}, {Dey}, {Liu}, \&
  {Graham}}]{2000ApJ...542...27I}
{Ivison}, R.~J., {Dunlop}, J.~S., {Smail}, I., {et~al.} 2000, \apj, 542, 27

\bibitem[{{Kauffmann} {et~al.}(1999){Kauffmann}, {Colberg}, {Diaferio}, \&
  {White}}]{1999MNRAS.307..529K}
{Kauffmann}, G., {Colberg}, J.~M., {Diaferio}, A., \& {White}, S.~D.~M. 1999,
  \mnras, 307, 529

\bibitem[{{Kennicutt}(1998)}]{1998ARA&A..36..189K}
{Kennicutt}, Jr., R.~C. 1998, \araa, 36, 189

\bibitem[{{Kubo} {et~al.}(2015){Kubo}, {Yamada}, {Ichikawa}, {Kajisawa},
  {Matsuda}, \& {Tanaka}}]{2015ApJ...799...38K}
{Kubo}, M., {Yamada}, T., {Ichikawa}, T., {et~al.} 2015, \apj, 799, 38

\bibitem[{{Lehmer} {et~al.}(2009{\natexlab{a}}){Lehmer}, {Alexander}, {Geach},
  {Smail}, {Basu-Zych}, {Bauer}, {Chapman}, {Matsuda}, {Scharf}, {Volonteri},
  \& {Yamada}}]{2009ApJ...691..687L}
{Lehmer}, B.~D., {Alexander}, D.~M., {Geach}, J.~E., {et~al.}
  2009{\natexlab{a}}, \apj, 691, 687

\bibitem[{{Lehmer} {et~al.}(2009{\natexlab{b}}){Lehmer}, {Alexander},
  {Chapman}, {Smail}, {Bauer}, {Brandt}, {Geach}, {Matsuda}, {Mullaney}, \&
  {Swinbank}}]{2009MNRAS.400..299L}
{Lehmer}, B.~D., {Alexander}, D.~M., {Chapman}, S.~C., {et~al.}
  2009{\natexlab{b}}, \mnras, 400, 299

\bibitem[{{Ma} {et~al.}(2015){Ma}, {Smail}, {Swinbank}, {Simpson}, {Thomson},
  {Chen}, {Danielson}, {Hilton}, {Tadaki}, {Stott}, \&
  {Kodama}}]{2015ApJ...806..257M}
{Ma}, C.-J., {Smail}, I., {Swinbank}, A.~M., {et~al.} 2015, \apj, 806, 257

\bibitem[{{Madau} \& {Dickinson}(2014)}]{2014ARA&A..52..415M}
{Madau}, P., \& {Dickinson}, M. 2014, \araa, 52, 415

\bibitem[{{Matsuda} {et~al.}(2004){Matsuda}, {Yamada}, {Hayashino}, {Tamura},
  {Yamauchi}, {Ajiki}, {Fujita}, {Murayama}, {Nagao}, {Ohta}, {Okamura},
  {Ouchi}, {Shimasaku}, {Shioya}, \& {Taniguchi}}]{2004AJ....128..569M}
{Matsuda}, Y., {Yamada}, T., {Hayashino}, T., {et~al.} 2004, \aj, 128, 569

\bibitem[{{Matsuda} {et~al.}(2005){Matsuda}, {Yamada}, {Hayashino}, {Tamura},
  {Yamauchi}, {Murayama}, {Nagao}, {Ohta}, {Okamura}, {Ouchi}, {Shimasaku},
  {Shioya}, \& {Taniguchi}}]{2005ApJ...634L.125M}
---. 2005, \apjl, 634, L125

\bibitem[{{Miller} {et~al.}(2015){Miller}, {Hayward}, {Chapman}, \&
  {Behroozi}}]{2015MNRAS.452..878M}
{Miller}, T.~B., {Hayward}, C.~C., {Chapman}, S.~C., \& {Behroozi}, P.~S. 2015,
  \mnras, 452, 878

\bibitem[{{Mullaney} {et~al.}(2011){Mullaney}, {Alexander}, {Goulding}, \&
  {Hickox}}]{2011MNRAS.414.1082M}
{Mullaney}, J.~R., {Alexander}, D.~M., {Goulding}, A.~D., \& {Hickox}, R.~C.
  2011, \mnras, 414, 1082

\bibitem[{{Scott} {et~al.}(2006){Scott}, {Dunlop}, \&
  {Serjeant}}]{2006MNRAS.370.1057S}
{Scott}, S.~E., {Dunlop}, J.~S., \& {Serjeant}, S. 2006, \mnras, 370, 1057

\bibitem[{{Silva} {et~al.}(1998){Silva}, {Granato}, {Bressan}, \&
  {Danese}}]{1998ApJ...509..103S}
{Silva}, L., {Granato}, G.~L., {Bressan}, A., \& {Danese}, L. 1998, \apj, 509,
  103

\bibitem[{{Simpson} {et~al.}(2014){Simpson}, {Swinbank}, {Smail}, {Alexander},
  {Brandt}, {Bertoldi}, {de Breuck}, {Chapman}, {Coppin}, {da Cunha},
  {Danielson}, {Dannerbauer}, {Greve}, {Hodge}, {Ivison}, {Karim}, {Knudsen},
  {Poggianti}, {Schinnerer}, {Thomson}, {Walter}, {Wardlow}, {Wei{\ss}}, \&
  {van der Werf}}]{2014ApJ...788..125S}
{Simpson}, J.~M., {Swinbank}, A.~M., {Smail}, I., {et~al.} 2014, \apj, 788, 125

\bibitem[{{Smail} {et~al.}(2003){Smail}, {Ivison}, {Gilbank}, {Dunlop}, {Keel},
  {Motohara}, \& {Stevens}}]{2003ApJ...583..551S}
{Smail}, I., {Ivison}, R.~J., {Gilbank}, D.~G., {et~al.} 2003, \apj, 583, 551

\bibitem[{{Smail} {et~al.}(2014){Smail}, {Geach}, {Swinbank}, {Tadaki},
  {Arumugam}, {Hartley}, {Almaini}, {Bremer}, {Chapin}, {Chapman}, {Danielson},
  {Edge}, {Scott}, {Simpson}, {Simpson}, {Conselice}, {Dunlop}, {Ivison},
  {Karim}, {Kodama}, {Mortlock}, {Robson}, {Roseboom}, {Thomson}, {van der
  Werf}, \& {Webb}}]{2014ApJ...782...19S}
{Smail}, I., {Geach}, J.~E., {Swinbank}, A.~M., {et~al.} 2014, \apj, 782, 19

\bibitem[{{Steidel} {et~al.}(1998){Steidel}, {Adelberger}, {Dickinson},
  {Giavalisco}, {Pettini}, \& {Kellogg}}]{1998ApJ...492..428S}
{Steidel}, C.~C., {Adelberger}, K.~L., {Dickinson}, M., {et~al.} 1998, \apj,
  492, 428

\bibitem[{{Swinbank} {et~al.}(2010){Swinbank}, {Smail}, {Longmore}, {Harris},
  {Baker}, {De Breuck}, {Richard}, {Edge}, {Ivison}, {Blundell}, {Coppin},
  {Cox}, {Gurwell}, {Hainline}, {Krips}, {Lundgren}, {Neri}, {Siana},
  {Siringo}, {Stark}, {Wilner}, \& {Younger}}]{2010Natur.464..733S}
{Swinbank}, A.~M., {Smail}, I., {Longmore}, S., {et~al.} 2010, \nat, 464, 733

\bibitem[{{Swinbank} {et~al.}(2014){Swinbank}, {Simpson}, {Smail}, {Harrison},
  {Hodge}, {Karim}, {Walter}, {Alexander}, {Brandt}, {de Breuck}, {da Cunha},
  {Chapman}, {Coppin}, {Danielson}, {Dannerbauer}, {Decarli}, {Greve},
  {Ivison}, {Knudsen}, {Lagos}, {Schinnerer}, {Thomson}, {Wardlow}, {Wei{\ss}},
  \& {van der Werf}}]{2014MNRAS.438.1267S}
{Swinbank}, A.~M., {Simpson}, J.~M., {Smail}, I., {et~al.} 2014, \mnras, 438,
  1267

\bibitem[{{Tamura} {et~al.}(2009){Tamura}, {Kohno}, {Nakanishi}, {Hatsukade},
  {Iono}, {Wilson}, {Yun}, {Takata}, {Matsuda}, {Tosaki}, {Ezawa}, {Perera},
  {Scott}, {Austermann}, {Hughes}, {Aretxaga}, {Chung}, {Oshima}, {Yamaguchi},
  {Tanaka}, \& {Kawabe}}]{2009Natur.459...61T}
{Tamura}, Y., {Kohno}, K., {Nakanishi}, K., {et~al.} 2009, \nat, 459, 61

\bibitem[{{Tamura} {et~al.}(2010){Tamura}, {Iono}, {Wilner}, {Kajisawa},
  {Uchimoto}, {Alexander}, {Chung}, {Ezawa}, {Hatsukade}, {Hayashino},
  {Hughes}, {Ichikawa}, {Ikarashi}, {Kawabe}, {Kohno}, {Lehmer}, {Matsuda},
  {Nakanishi}, {Takata}, {Wilson}, {Yamada}, \& {Yun}}]{2010ApJ...724.1270T}
{Tamura}, Y., {Iono}, D., {Wilner}, D.~J., {et~al.} 2010, \apj, 724, 1270

\bibitem[{{Taniguchi} \& {Shioya}(2000)}]{2000ApJ...532L..13T}
{Taniguchi}, Y., \& {Shioya}, Y. 2000, \apjl, 532, L13

\bibitem[{{Uchimoto} {et~al.}(2012){Uchimoto}, {Yamada}, {Kajisawa}, {Kubo},
  {Ichikawa}, {Matsuda}, {Akiyama}, {Hayashino}, {Konishi}, {Nishimura},
  {Omata}, {Suzuki}, {Tanaka}, {Tokoku}, \& {Yoshikawa}}]{2012ApJ...750..116U}
{Uchimoto}, Y.~K., {Yamada}, T., {Kajisawa}, M., {et~al.} 2012, \apj, 750, 116

\bibitem[{{Umehata} {et~al.}(2014){Umehata}, {Tamura}, {Kohno}, {Hatsukade},
  {Scott}, {Kubo}, {Yamada}, {Ivison}, {Cybulski}, {Aretxaga}, {Austermann},
  {Hughes}, {Ezawa}, {Hayashino}, {Ikarashi}, {Iono}, {Kawabe}, {Matsuda},
  {Matsuo}, {Nakanishi}, {Oshima}, {Perera}, {Takata}, {Wilson}, \&
  {Yun}}]{2014MNRAS.440.3462U}
{Umehata}, H., {Tamura}, Y., {Kohno}, K., {et~al.} 2014, \mnras, 440, 3462

\bibitem[{{Walter} {et~al.}(2012){Walter}, {Decarli}, {Carilli}, {Bertoldi},
  {Cox}, {da Cunha}, {Daddi}, {Dickinson}, {Downes}, {Elbaz}, {Ellis}, {Hodge},
  {Neri}, {Riechers}, {Weiss}, {Bell}, {Dannerbauer}, {Krips}, {Krumholz},
  {Lentati}, {Maiolino}, {Menten}, {Rix}, {Robertson}, {Spinrad}, {Stark}, \&
  {Stern}}]{2012Natur.486..233W}
{Walter}, F., {Decarli}, R., {Carilli}, C., {et~al.} 2012, \nat, 486, 233

\bibitem[{{Wang} {et~al.}(2013){Wang}, {Brandt}, {Luo}, {Smail}, {Alexander},
  {Danielson}, {Hodge}, {Karim}, {Lehmer}, {Simpson}, {Swinbank}, {Walter},
  {Wardlow}, {Xue}, {Chapman}, {Coppin}, {Dannerbauer}, {De Breuck}, {Menten},
  \& {van der Werf}}]{2013ApJ...778..179W}
{Wang}, S.~X., {Brandt}, W.~N., {Luo}, B., {et~al.} 2013, \apj, 778, 179

\bibitem[{{Webb} {et~al.}(2009){Webb}, {Yamada}, {Huang}, {Ashby}, {Matsuda},
  {Egami}, {Gonzalez}, \& {Hayashimo}}]{2009ApJ...692.1561W}
{Webb}, T.~M.~A., {Yamada}, T., {Huang}, J.-S., {et~al.} 2009, \apj, 692, 1561

\bibitem[{{Wei{\ss}} {et~al.}(2013){Wei{\ss}}, {De Breuck}, {Marrone},
  {Vieira}, {Aguirre}, {Aird}, {Aravena}, {Ashby}, {Bayliss}, {Benson},
  {B{\'e}thermin}, {Biggs}, {Bleem}, {Bock}, {Bothwell}, {Bradford}, {Brodwin},
  {Carlstrom}, {Chang}, {Chapman}, {Crawford}, {Crites}, {de Haan}, {Dobbs},
  {Downes}, {Fassnacht}, {George}, {Gladders}, {Gonzalez}, {Greve},
  {Halverson}, {Hezaveh}, {High}, {Holder}, {Holzapfel}, {Hoover}, {Hrubes},
  {Husband}, {Keisler}, {Lee}, {Leitch}, {Lueker}, {Luong-Van}, {Malkan},
  {McIntyre}, {McMahon}, {Mehl}, {Menten}, {Meyer}, {Murphy}, {Padin},
  {Plagge}, {Reichardt}, {Rest}, {Rosenman}, {Ruel}, {Ruhl}, {Schaffer},
  {Shirokoff}, {Spilker}, {Stalder}, {Staniszewski}, {Stark}, {Story},
  {Vanderlinde}, {Welikala}, \& {Williamson}}]{2013ApJ...767...88W}
{Wei{\ss}}, A., {De Breuck}, C., {Marrone}, D.~P., {et~al.} 2013, \apj, 767, 88

\bibitem[{{Yamada} {et~al.}(2012){Yamada}, {Nakamura}, {Matsuda}, {Hayashino},
  {Yamauchi}, {Morimoto}, {Kousai}, \& {Umemura}}]{2012AJ....143...79Y}
{Yamada}, T., {Nakamura}, Y., {Matsuda}, Y., {et~al.} 2012, \aj, 143, 79

\end{thebibliography}
\end{document}